\documentclass[a4paper,12pt]{article}
\usepackage{epsfig}
\usepackage{a4wide}
\usepackage[latin1]{inputenc}
\usepackage{dsfont}
\usepackage{amsfonts}
\usepackage{amsmath}
\usepackage{amssymb}
\usepackage{amsthm}
\usepackage{graphicx}
\usepackage{booktabs}
\usepackage[numbers, sort]{natbib}
\allowdisplaybreaks[1]
\usepackage{color}
\usepackage{ulem}

\usepackage{epstopdf}

\usepackage[small,bf]{caption}
\newcommand{\be}{\begin{equation}}
\newcommand{\ee}{\end{equation}}
\newcommand{\beq}{\begin{equation}}
\newcommand{\eeq}{\end{equation}}
\newcommand{\bea}{\begin{eqnarray}}
\newcommand{\eea}{\end{eqnarray}}

\newcommand{\braket}[1]{\langle #1 \rangle}
\newcommand{\pd}[2]{\frac{ \partial #1 }{ \partial #2 }}
\newcommand{\pdd}[2]{\frac{ \partial^2 #1 }{ \partial #2^2 }}

\begin{document}

\begin{titlepage}

\vspace*{-15mm}
\vspace*{0.7cm}

\begin{center}

{\Large {\bf Pseudosmooth Tribrid Inflation}}\\[8mm]

Stefan Antusch$^{\star\dagger}$\footnote{Email: \texttt{stefan.antusch@unibas.ch}},~
David Nolde$^{\star}$\footnote{Email: \texttt{david.nolde@unibas.ch}}, and
Mansoor Ur Rehman$^{\star}$\footnote{Email: \texttt{mansoor-ur.rehman@unibas.ch}}~

\end{center}

\vspace*{0.20cm}

\centerline{$^{\star}$ \it
Department of Physics, University of Basel,}
\centerline{\it
Klingelbergstr.\ 82, CH-4056 Basel, Switzerland}

\vspace*{0.4cm}

\centerline{$^{\dagger}$ \it
Max-Planck-Institut f\"ur Physik (Werner-Heisenberg-Institut),}
\centerline{\it
F\"ohringer Ring 6, D-80805 M\"unchen, Germany}

\vspace*{1.2cm}

\begin{abstract}
\noindent
We explore a new class of supersymmetric models of inflation where the inflaton is realised as a combination of a Higgs field and (gauge non-singlet) matter fields, using a ``tribrid''  structure of the superpotential.  Inflation is associated with a phase transition around GUT scale energies. The inflationary trajectory already preselects the later vacuum after inflation, which has the advantage of automatically avoiding the production of dangerous topological defects at the end of inflation. While at first sight the models look similar to smooth inflation, they feature a waterfall and are therefore only pseudosmooth. The new class of models offers novel possibilities for realising inflation in close contact with particle physics, for instance with supersymmetric GUTs or with supersymmetric flavour models based on family symmetries.
\end{abstract}

\end{titlepage}

\section{Introduction}
Inflation provides a successful paradigm for solving the horizon and flatness problems of the standard Big Bang cosmology, and at the same time for explaining the origin of structure of the observable Universe~\cite{Guth:1980zm,Liddle:2000cg}. Several schemes for inflation have been proposed including chaotic inflation~\cite{Linde:1983gd}, which
predicts large tensor perturbations~\cite{Lyth:1996im}, in contrast to hybrid inflation~\cite{Linde:1991km} which predicts small ones. While in chaotic inflation the field values during inflation exceed the Planck scale, they can be well below the Planck scale in hybrid inflation. This allows for an effective field theory (EFT) treatment, in particular for a small field expansion of the K\"ahler potential in the effective supergravity (SUGRA) theory. Hybrid inflation is furthermore connected to a phase transition in particle physics, which may be identified as, for instance, the spontaneous breaking of a Grand Unified Theory (GUT) group \cite{Dvali:1994ms} or of a family symmetry \cite{Antusch:2008gw}. This, in principle, offers the possibility for a close connection between particle theories and inflationary cosmology, see for example \cite{Senoguz:2003zw}-\cite{Kyae:2005nv}.

There are two ways to implement the hybrid inflation scheme in supersymmetric (SUSY) theories. In the most common approaches, such as the ``standard'' SUSY hybrid inflation models \cite{Copeland:1994vg,Dvali:1994ms,Linde:1997sj,Senoguz:2004vu,BasteroGil:2006cm} or in D-term hybrid inflation \cite{Binetruy:1996xj,Halyo:1996pp}, the inflaton is a gauge singlet field. For instance, in ``standard'' SUSY hybrid inflation the inflaton is identified as one of the ``driving fields'' which generate the potential for the waterfall field by its F-term, but are otherwise disconnected from the rest of the theory. Although inflation ends by a particle physics phase transition, the connection to the particle theory is therefore still rather loose.

An alternative possibility is ``tribrid inflation'' \cite{Antusch:2004hd,Antusch:2008pn,Antusch:2009vg,Antusch:2010va}, where the inflaton can be a (gauge non-singlet) matter field. Models of this type are referred to as tribrid inflation since they employ three fields for realising inflation: The inflaton field direction in the matter sector, plus a waterfall field and a driving field as in conventional SUSY hybrid inflation. Since the inflaton forms part of the matter sector, there can be close connections to particle physics models. For example, the inflaton can be a D-flat direction of fields in GUT representations in SO(10) GUTs or Pati-Salam models \cite{Antusch:2010va}. Another attractive possibility is that the inflaton in tribrid inflation can be identified as one of the right-handed sneutrinos \cite{Antusch:2004hd,Antusch:2010mv}.\footnote{The possibility that the right-handed sneutrino may act as the inflaton has first been proposed in the context of chaotic inflation in \cite{Murayama:1992ua}.} Furthermore, in tribrid inflation it has been shown that the $\eta$-problem can be solved, in the context of supergravity, by either a small field expansion of the K\"ahler potential (with adjustment of expansion parameters) but also by a Heisenberg symmetry \cite{Antusch:2008pn,Antusch:2011ei} or a shift symmetry \cite{Antusch:2009ef} in the K\"ahler potential.    

When connecting inflation to a particle physics phase transition, one has to take care that dangerous topological defects, like monopoles, are not produced abundantly after inflation. In tribrid inflation with a gauge non-singlet inflaton, monopole production may be avoided since the symmetry (for instance the GUT symmetry) is already broken during inflation by the inflaton vacuum expectation value \cite{Antusch:2010va}. In the context of hybrid inflation with a gauge-singlet inflaton, known strategies are ``smooth'' \cite{smooth,Rehman:2012gd} inflation or ``shifted'' \cite{Jeannerot:2000sv,Civiletti:2011qg} inflation. In these models, the inflationary potential is deformed such that the waterfall field is already displaced from zero during inflation. In all solutions, the key is to enforce that the waterfall which ends inflation happens in one ``preselected'' direction in field space. When this is the case, no topological defects can form.

In this paper, we therefore investigate how the production of topological defects can be avoided in variants of tribrid inflation with ``vacuum preselection''. We find new classes of tribrid inflation models where the waterfall field is non-zero during inflation. The inflaton direction is now a combination of a Higgs field and (gauge non-singlet) matter fields. The inflationary trajectory already preselects the later vacuum after inflation, which has the desired effect of avoiding the production of dangerous topological defects at the end of inflation. While at first sight the field dynamics looks similar to smooth inflation, it turns out that the tribrid models feature a waterfall - in contrast to smooth inflation where the the field trajectory is always in a minimum of the potential. So one may say that the dynamics is only ``pseudosmooth''. Regarding the inflationary predictions, the WMAP best-fit values can be accommodated with field values well below the Planck scale and the phase transition is found to happen around GUT scale energies.

The paper is organised as follows: In Section 2, we consider a generalized form of tribrid inflation models
and derive necessary conditions for the inflationary trajectory with a non-zero value of Higgs field $(H  \neq 0)$. 
In order to demonstrate its qualitative features, we provide an approximate analytical treatment of the
model in Section 3. Here, we not only derive the appearance of a waterfall, we also calculate the predictions of various 
cosmic microwave background (CMB) observables. Our numerical results are discussed in Section 4 for a specific case, and are found
compatible with the approximate analytical estimates. Finally, we conclude with a summary of our results in Section 5.

\section{A new class of models from generalized tribrid \\ structure}
We consider a superpotential with generalized tribrid structure:
\be
\label{eqW}
W = \kappa \left\{  S( H^l - M^2 ) + \lambda H^m \phi^n  \right\}.
\ee
The SUSY F-term potential then is
\be
\label{eqVF}
V_F \equiv \sum_i \left| \frac{\partial W}{\partial z_i} \right|
=\kappa^2 \left\{  \left|  H^l - M^2  \right|^2  +  \left|  l S H^{l-1} + m \lambda H^{m-1} \phi^n  \right|^2  +  \left|  n \lambda H^m \phi^{n-1}  \right|^2  \right\},
\ee
with $z_i = \{S,\,H,\,\phi \}$, and we use the same notation for superfields and their scalar components.
Here and below we use Planck units $m_P = 1$, where $m_P = 2.4 \times 10^{18}$ GeV is the reduced Planck mass.

The scalar potential can get SUGRA corrections from the K\"ahler potential, which generate a mass term $\kappa_S \kappa^2 M^4 |S|^2$ for the driving field $S$. This SUGRA mass is required to suppress $\braket{S}$ during inflation, because a large $\braket{S}$ can make the effective inflaton potential too steep. Such an extra mass term can be generated e.g.~by a K\"ahler potential term $\Delta K = -\frac14 \kappa_S (S^\dagger S)^2$ with $\kappa_S = O(1)$. Apart from the extra mass term for $S$, we will use the potential \eqref{eqVF} to keep the discussion simple. More complicated K\"ahler potentials can be used, but the purpose of this paper is to demonstrate that this new model class contains phenomenologically viable models, not to provide a full description of all possibilities.

\begin{figure}[t]
\centering \includegraphics[width=12 cm]{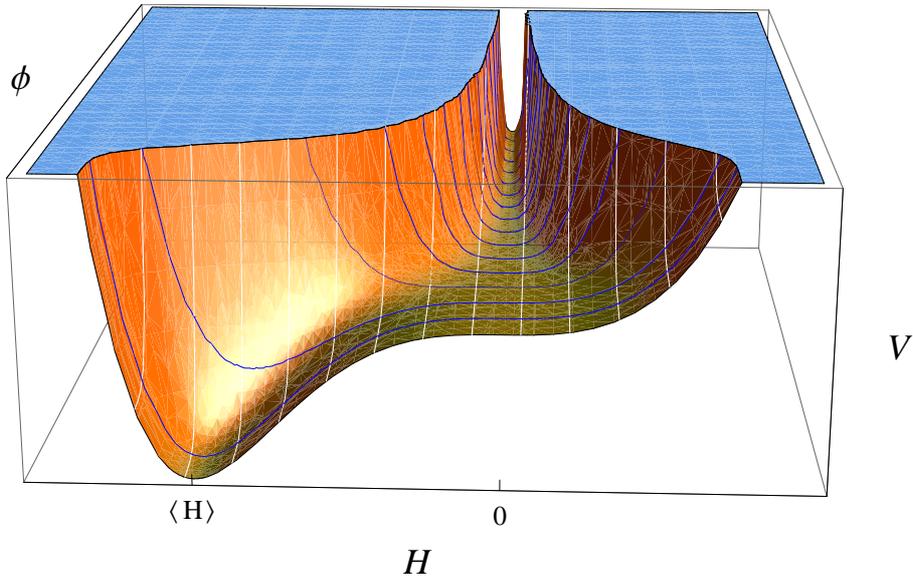}
\caption{The scalar potential $V$ as a function of $H$ and $\phi$ with $S = S_{\text{min}}$ and $l=m=3,\,n=2$.
The inflationary valley in the figure is slightly shifted from $H=0$, which can be seen more clearly in Fig.~\ref{fig2}.}
\label{fig1}
\end{figure}

The parameter $\kappa$ can be used to scale the potential to generate the desired amplitude of curvature perturbations of $\Delta_{\mathcal{R}}^2(k_0) = (2.43 \pm 0.11) \times 10^{-9}$ at the pivot scale $k_0 = 0.002 \text{ Mpc}^{-1}$ \cite{Komatsu:2010fb}. It has no other effect and will therefore be ignored throughout this paper. We thus work with the SUSY potential \eqref{eqVF}, corrected by a SUGRA mass term for $S$, and suppressing the amplitude parameter $\kappa$:
\be
V = \left|  H^l - M^2  \right|^2  +  \left|  l S H^{l-1} + m \lambda^2 H^{m-1} \phi^n  \right|^2  +  \left|  n \lambda^2 H^m \phi^{n-1}  \right|^2  +  \kappa_S M^4 \left| S \right|^2.
\ee
The potential for the example case of $l=m=3$, $n=2$ with $S$ in its minimum is shown in Fig.~\ref{fig1}.

\subsection{The role of $\phi$, $H$ and $S$ during inflation}

$\phi$ is a light field (for $H=S=0$ its potential is flat at tree level) which is slowly-rolling during inflation. $H$ and $S$ are heavy fields which stay near their minima during inflation. The positions of the minima depend on $\phi$ - which is changing over time - so they slowly move during inflation as well, tracking their moving minima.

Inflation therefore happens along a multi-field trajectory where $\phi$, $H$ and $S$ are all non-zero. We can usually assume that mostly $\phi$ is moving and $H$ and $S$ are very small during inflation, so that we have an effective single-field model with $\phi$ as the inflaton. The small non-zero $\braket{H}$ and $\braket{S}$ then induce small potential terms for the inflaton $\phi$, generating the slope which allows the $\phi$ field to slowly roll towards its minimum at $\phi = 0$.

\begin{figure}[t]
\centering \includegraphics[width=12cm]{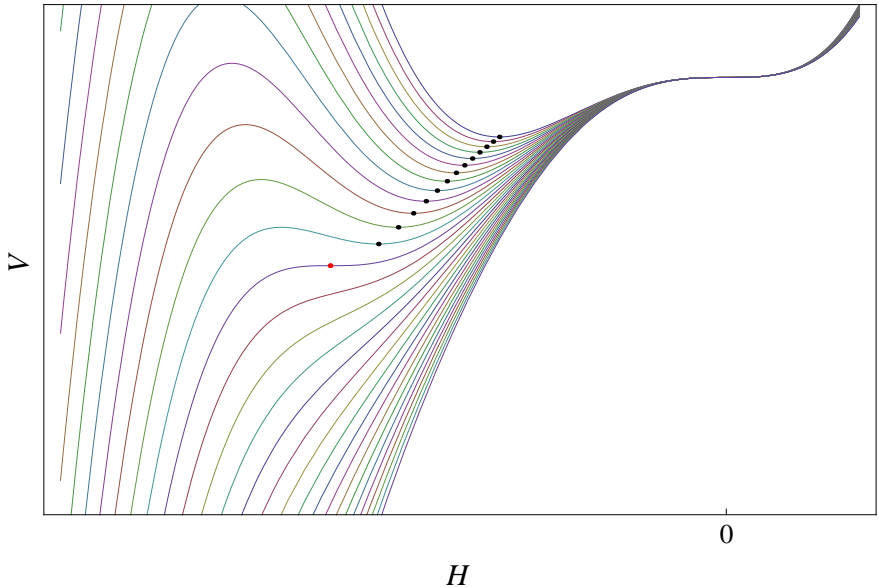}
\caption{The scalar potential $V$ as a function of $H$ with $l=m=3$ and $n=2$, for various values of $\phi$ 
near the waterfall which starts at the red dot. The small black dots represent minima of the potential along the inflationary track.}
\label{fig2}
\end{figure}

So far the model looks like a tribrid model of smooth inflation. More careful analysis reveals a crucial difference, though: These tribrid models end with a waterfall (see Fig.~\ref{fig2}). The non-zero value of $\braket{S}$ induces a negative mass term for the $H$ field, and below some critical inflaton value $\phi_c$ this negative mass term becomes dominant and triggers a waterfall transition of the $H$ field.

The proposed model therefore combines properties of smooth inflation, e.g.~a non-vanishing $\braket{H}$ during inflation which prevents the formation of topological defects, and properties of standard hybrid inflation, e.g.~a waterfall phase transition which could produce peculiar features in the spectrum of primordial perturbations \cite{Lyth:2012yp}. We also retain the tribrid structure's benefit that the inflaton $\phi$ can be a gauge-non-singlet, which makes it easier to have inflation in motivated particle physics models, e.g.~GUT or flavour models.

\subsection{Allowed superpotential parameters $l$ and $m$}

An order-of-magnitude estimate demonstrates that these models require $l \geq m > 2$ (if $\braket{S}$ is negligible), if we require the fields to be sub-Planckian.

\begin{table}[t]
\centering
  \begin{tabular}{ | l | c c c c | }
    \hline
  & $m=2$ & $m=3$ & $m = 4$ & $m=5$ \\
    \hline
  $l=2$ & \textcolor{red}{$m \leq 2$} & \textcolor{red}{$m > l$} & \textcolor{red}{$m > l$} & \textcolor{red}{$m > l$}  \\
  $l=3$ & \textcolor{red}{$m \leq 2$} & \textcolor[rgb]{0,0.7,0}{possible} & \textcolor{red}{$m > l$} & \textcolor{red}{$m > l$}  \\
  $l=4$ & \textcolor{red}{$m \leq 2$} & \textcolor[rgb]{1,0.5,0}{with $m_{H,SUGRA}$} & \textcolor[rgb]{0,0.7,0}{possible} & \textcolor{red}{$m > l$}  \\
  $l=5$ & \textcolor{red}{$m \leq 2$} & \textcolor[rgb]{1,0.5,0}{with $m_{H,SUGRA}$} & \textcolor[rgb]{1,0.5,0}{possible} & \textcolor[rgb]{0,0.7,0}{possible} \\
    \hline
  \end{tabular}
\caption{Viable (green) and dysfunctional (red) superpotential choices for the proposed model class. The red text indicates why condition \eqref{mlCondition} is violated. The orange entries satisfy the necessary condition \eqref{mlCondition} and may be smooth or pseudosmooth, but are not studied in detail in this paper. Some of these models can only work with additional SUGRA contributions to $m_H^2$. Larger $l, m$ are possible, but increasingly Planck-suppressed. Superpotentials which do not work for this model class can still provide inflation by other mechanisms (e.g.~$l\geq m=2$ for conventional tribrid inflation \cite{Antusch:2004hd,Antusch:2008pn,Antusch:2009vg,Antusch:2010va}).}
\label{tab:summaryML}
\end{table}

For $S \simeq 0$, $H \ll \phi$ and $H^l \ll M^2$, the potential simplifies to\footnote{$S \simeq 0$ works well for sufficiently large $\kappa_S$. $\braket{H^l} = M^2$ is the $H$ vev after inflation; $H^l$ will be much smaller before the waterfall, and $H \ll \phi$ is a condition we impose because we are looking for tribrid-like models where the inflaton is composed mostly of the $\phi$ field.}
\be
V  \simeq  M^4  -  2M^2 H^l  +  m^2 \lambda^2 H^{2m-2} \phi^{2n} .
\ee
We now minimise this potential with respect to $H$ to find the field $H_{\text{min}}(\phi)$. The result is
\be
M^{-2} H_{\text{min}}^{2m-l-2}  \simeq  \frac{l}{m^2(m-1)} \frac{1}{\lambda^2 \phi^{2n}} .
\ee
On the left-hand side we use that $H^{-l} \gg M^{-2}$ and on the right-hand side we use $\phi^{2n} < 1$:
\be
H_{\text{min}}^{2m-2l-2}  \gg  M^{-2} H_{\text{min}}^{2m-l-2}  \simeq  \frac{l}{m^2(m-1)} \frac{1}{\lambda^2 \phi^{2n}}  \gtrsim  O(1)
\ee
\be
\Rightarrow \quad 2m - 2l - 2  <  0.
\ee
This demonstrates that $2m - 2l - 2  <  0$ is required to simultaneously achieve $\phi < 1$ (sub-Planckian fields) and $H_{\text{min}} \ll \phi$ (tribrid-like inflation with $\phi$ as the main inflaton component).

To get combined Higgs-matter inflation instead of conventional tribrid inflation with $H = 0$, it is also required that $H = 0$ must not be a stable minimum. In particular, the $m^2 \lambda^2 \phi^{2n} H^{2m-2}$ term must not be the one with the lowest power of $H$. Without further SUGRA correction terms, this implies that $-2M^2 H^l$ needs to have a smaller exponent, so $l < 2m-2$. If we allow an extra SUGRA mass term for $H$, with negative coefficient, this condition is relaxed to $2 < 2m-2$, or equivalently $m > 2$.

For integer values of $l$ and $m$, this leads to the combined constraints:
\begin{align}
 l \geq m &>  1 + \frac{l}{2}   & \text{(without SUGRA mass term for $H$)} \notag \\
 \label{mlCondition}
 l \geq m & > 2  & \text{(if negative SUGRA mass term for $H$ is present)}.
\end{align}
In both cases the constraints imply that $l, m > 2$. Table \ref{tab:summaryML} contains an overview of the possibilities for $l$ and $m$.

We see that the proposed models can account for (small-field) inflation for $l = m \geq 3$. Models with $l > m$ may also be possible, but either depend more strongly on the K\"ahler potential (like for $l=4,\,m=3$) or require more strongly Planck-suppressed operators (like for $l=5,\,m=4$). In the following sections, we will analyse the models with $l=m$ in more detail, and leave models with $l > m$ for later study.

\section{Approximate analytical treatment for $l=m$}
In this section we want to find analytic estimates for the predictions of the considered model class for $l=m$. We will first demonstrate some of its qualitative features, including the tachyonic instability leading to the waterfall. We will then continue to determine the predicted CMB spectrum and its dependence on the model parameters $M$ and $\lambda$.

Unfortunately, the non-trivial dynamics of the $H$ and $S$ fields during inflation can make exact computations very complicated. To arrive at comprehensible results, we will resort to several approximations. We will assume that $H$ and $S$ track their $\phi$-dependent minima perfectly, and for most purposes we will set $\braket{S} \simeq 0$ entirely. Particularly the latter approximation may not be perfectly accurate, but it is good enough for the purpose of this section. For more precise results, we refer to the numerical analysis in the next section.

We will also assume that the phases of the fields are chosen to minimise the potential, and we write $\phi \equiv \left| \phi \right|$, $H \equiv \left| H \right|$ and $S \equiv \left| S \right|$.

\subsection{Determination of $S_{\text{min}}$}
We will usually assume that $S$, being much heavier than $\phi$, perfectly tracks its minimum. We will now determine $S_{\text{min}}(H, \phi)$ and replace the dynamical field $S$ with $S_{\text{min}}$, reducing the effective number of dynamical fields to $H$ and $\phi$.

$S_{\text{min}}$ is determined by minimizing the potential 
\be
\pd{V}{S} = 2\left(  l^2 H^{2l-2} + \kappa_S M^4  \right)  S  -  2 l^2\lambda H^{2l-2} \phi^n  \stackrel{!}{=}  0
\ee
and solving for
\be
S_{\text{min}} = \frac{ l^2 \lambda \phi^n H^{2l-2} }{ \kappa_S M^4 + l^2 H^{2l-2} } .
\ee
If we plug this back into our potential we get the effective 2-field potential
\begin{align}
V_{2} &= V(\phi, H, S_{\text{min}}(\phi, H))\\
&=M^4 - 2M^2 H^l + H^{2l} + n^2 \lambda^2 H^{2l} \phi^{2n-2} + l^2 \lambda^2 H^{2l-2} \phi^{2n} - \frac{ l^4 \lambda^2 \phi^{2n} H^{4l-4} }{ \kappa_S M^4 + l^2 H^{2l-2} } \\
&\simeq M^4 - 2M^2 H^l + l^2 \lambda^2 H^{2l-2} \phi^{2n} - \frac{ l^4 \lambda^2 \phi^{2n} H^{4l-4} }{ \kappa_S M^4 + l^2 H^{2l-2} },
\end{align}
where in the last line we have used that $H^l \ll M^2$ (because $\braket{H^l} = M^2$ after inflation, and $H$ is much smaller before the waterfall) and $H \ll \phi$ during inflation.

\subsection{Waterfall at $\phi_c$}
In this section, we want to find the critical inflaton value $\phi_c$ where $H$ exhibits a tachyonic instability. The critical point where the waterfall starts is characterized by two vanishing derivatives:
\be
\pd{V_2}{H}(H_c, \phi_c) = \pdd{V_2}{H}(H_c, \phi_c) = 0 .
\ee
The first derivative is zero because up to this point, $H$ is assumed to track its minimum, and the second derivative is zero because $m_H^2$ changes its sign.\footnote{More precisely, one should look for a vanishing eigenvalue in the $2 \times 2$ mass matrix involving $H$ and $S$. The off-diagonal elements are quite small though, so we neglect them for simplicity.} These conditions can be solved for $\phi_c$ and $H_c$ with the result

\begin{align}
\label{phic}
\phi_c^{2n} &= \frac{ 16(l-1) }{ l(3l-2)^2 } \frac{ M^2 }{ \lambda^2 } \left(  \frac{ 3l-2 }{ l-2 } \frac{ l^2 }{ \kappa_S M^4 }  \right)^{ \frac{l-2}{2l-2} } ,\\
H_c^{2l-2} &= \left(  \frac{ l-2 }{ 3l-2 } \frac{ \kappa_S M^4 }{ l^2 }  \right) .
\end{align}

This instability is a special feature of the model which sets it apart from smooth inflation. The dominant negative contribution to $m_H^2$ is not the $-M^2 H^l$ term of the potential, but rather the $\braket{S}$-induced term which becomes important near the critical $\phi_c$. This can also be seen by the fact that for $\kappa_S \rightarrow \infty$, which implies $\braket{S} \rightarrow 0$, the critical point is driven to $\phi_c \rightarrow 0$ (so there is no region with $\phi < \phi_c$, which means that there is no tachyonic instability).

\subsection{Effective inflaton potential $V_1$}
The following calculations can be very much simplified if we first neglect the $\braket{S}$-induced term in $V_2$. We are left with
\be
V_2 \simeq M^4 - 2M^2 H^l + l^2 \lambda^2 H^{2l-2} \phi^{2n} .
\ee

Near $\phi = \phi_0$, where the predictions for the CMB spectrum are calculated\footnote{$\phi_0$ is the value of the inflaton field $N_0 \simeq 50-60$ e-folds before the end of inflation, where perturbations with wavenumber $k_0 = 0.002~\text{Mpc}^{-1}$ leave the horizon.}, this approximation is good. Therefore it should be suitable for an estimate of the CMB spectrum. To reduce $V_2$ to the single-field potential $V_1$, we first determine the minimum of $V_2$ with respect to $H$:
\be
\pd{V_2}{H} = - 2l M^2 H^{l-1} + l^2 (2l-2) \lambda^2 H^{2l-3} \phi^{2n}  \stackrel{!}{=}  0 ,
\ee
which can easily be solved for $H$ to find
\be
H_{\text{min}}^{l-2} = \frac{ 1 }{ l(l-1) } \frac{ M^2 }{ \lambda^2 \phi^{2n} } .
\ee
We can insert this in $V_2$ to get the effective single-field potential
\begin{align}
V_1 &= V_2(\phi, H_{\text{min}}(\phi)) \\
&= M^4 - M^2 \frac{l-2}{l-1} \left(  \frac{ 1 }{ l(l-1) } \frac{ M^2 }{ \lambda^2 }  \right)^{ \frac{l}{l-2} }  \left(  \frac{1}{\phi}  \right)^{ \frac{2nl}{l-2} } .
\end{align}
We can now use this simple potential to derive the slow-roll predictions.

\subsection{CMB spectrum predictions}

\subsubsection{Slow-roll parameters}

The form of $V_1$ implies that $\varepsilon \ll \eta^2$ and $\xi \simeq \eta^2$, so $\eta$ will be the only relevant slow-roll parameter:
\begin{align}
V_1' &= V_0 C  \left(  \frac{1}{\phi}  \right)^{ 1 + \frac{2nl}{l-2} } \\
V_1'' &= -V_0 C \left(  1 + \frac{2nl}{l-2}  \right)  \left(  \frac{1}{\phi}  \right)^{ 2 + \frac{2nl}{l-2} } \\
&= -\left(  1 + \frac{2nl}{l-2}  \right)  \frac{1}{\phi}  V_1',
\end{align}
where
\be
C =  \frac{2nl}{l-1}  \frac{1}{M^2}  \left(  \frac{ 1 }{ l(l-1) } \frac{ M^2 }{ \lambda^2 }  \right)^{ \frac{l}{l-2} } .
\ee
This implies, using $\left(  1 + \frac{2nl}{l-2}  \right) > 3$ and $\phi \lesssim \frac13$, that
\be
\left| V_1' \right| \lesssim \frac{1}{10} \left| V_1'' \right| .
\ee

Using the definitions of the slow-roll parameters (with a factor of $\frac{1}{\sqrt{2}}$ for each derivative due to normalization of the fields), we find that
\be
\label{epsilonEta}
\varepsilon = \frac{1}{4} \left( \frac{V_1'}{V_1} \right)^2
  \lesssim \frac{1}{100} \left( \frac{1}{2} \frac{V_1''}{V_1} \right)^2
  =  \frac{\eta^2}{100}
\ee
and
\be
\label{xiEta}
\xi = \frac{1}{4} \frac{V_1' V_1'''}{V_1^2}
  = \frac{1}{4}  \left(  \frac{ 2 + \frac{2nl}{l-2} }{ 1 + \frac{2nl}{l-2} }  \right)   \left(  \frac{V_1''}{V_1}  \right)^2
  \simeq \frac{1}{4}  \left(  \frac{V_1''}{V_1}  \right)^2 
  = \eta^2 .
\ee
This immediately implies small $r$ and $\alpha_s$, because
\begin{align}
&(1 - n_s) = 6\,\varepsilon_0 - 2\,\eta_0 \simeq -2\,\eta_0 \\
\Rightarrow \quad  &| \eta_0 | < \frac{1}{20} \quad \text{(from WMAP constraints)}, \\
&r =  16\,\varepsilon_0 \lesssim \frac{16}{100} \eta_0^2 < \frac{1}{4000}, \\
&\left| \alpha_s \right| = \left| -2\,\xi_0 + 16\,\varepsilon_0 \,\eta_0 - 24\,\varepsilon_0^2 \right|  \simeq 2\,\xi_0^2  \simeq 2\,\eta_0^2 < \frac1{200},
\end{align}
where $\varepsilon_0$, $\eta_0$ and $\xi_0$ are the slow-roll parameters evaluated at $\phi_0$.

We will show below that $\eta$ and therefore also $r$ and $\alpha_s$ are even smaller, but this simple calculation at $\phi_0$ is sufficient to see that $r$ and $\alpha_s$ are too small to be measurable with the current experimental sensitivity.\footnote{The advantage of only using slow-roll parameters at $\phi_0$ is that we did not need to include the dynamics near $\phi_c$ for this argument. Therefore neglecting $\braket{S}$ is a very good approximation, and the result is quite robust.}

\subsubsection{Determination of $\phi_0$}

We now need to determine $\phi_0$ to find the spectral index. It can be deduced from the number of e-folds $N_0 \gtrsim 50$ which is given by
\begin{align}
&N_0  =  \int\limits_{\phi_c}^{\phi_0} \frac{ 2 V_1 }{ V_1' } ~\text{d}\phi   \simeq   \frac{2}{C} \int\limits_{\phi_c}^{\phi_0}   \phi^{1 + \frac{2nl}{l-2}}   ~\text{d}\phi\\
\label{phio}
\Rightarrow \quad  &\phi_0^{2 + \frac{2nl}{l-2}}  \simeq  \phi_c^{2 + \frac{2nl}{l-2}}  +  \frac{C}{2} \left(  2 + \frac{2nl}{l-2}  \right) N_0 ,
\end{align}
where the factor of $2$ in the first equation appeared due to the normalization of the inflaton field.

Eq.~\eqref{phio} was derived assuming that inflation ends due to the waterfall transition. It is also possible that the slow-roll conditions are violated before the waterfall happens. In that case, inflation does not end at $\phi_c$, but at $\phi_{\text{end}} > \phi_c$ where $| \eta( \phi_{\text{end}} ) |  = 1$. One can show, however, that this does not significantly change eq.~\eqref{phio}. Solving $| \eta( \phi_{\text{end}} ) | = 1$ for $\phi_{\text{end}}$, we find
\be
\phi_{\text{end}}^{2 + \frac{2nl}{l-2}} \simeq \frac{C}{2} \left( 1 + \frac{ 2nl }{ l-2 }  \right)  \ll  \frac{C}{2} \left(  2 + \frac{ 2nl }{ l-2 }  \right) N_0
\ee
which is much smaller (by a factor of $O(N_0)$) than the other term contributing to $\phi_0$ in eq.~\eqref{phio}. It can therefore be neglected in that equation. As $\phi_c \leq \phi_{\text{end}}$, there is no harm in replacing the (negligibly small) $\phi_{\text{end}}$ with the (also negligible) $\phi_c$. We thus see that $\phi_0$ as given in \eqref{phio} is a good approximation not only if inflation ends due to the waterfall at $\phi_c$, but even if inflation ends prematurely by a violation of the slow-roll conditions.

\subsubsection{Predicted CMB spectrum}
Inserting $\phi_0$ from eq.~\eqref{phio} to find $\eta_0$ and plugging in $\phi_c$ from eq.~\eqref{phic} leads to the final result for the slow-roll parameter $\eta_0$:
\be
\label{eta0}
\eta_0  =  - \left(  \frac{ 1 + \frac{2nl}{l-2} }{ 2 + \frac{2nl}{l-2} }  \right)   \frac{ 1 }{ N_0  +  c_{ln} \left(  \frac1M  \right)^{\frac{2}{l-1}(1 - \frac1n)}   \left(  \frac1{\kappa_S}  \right)^{\frac{l}{2l-2}(1 + \frac{l-2}{nl}) }    \left(  \frac1{\lambda}  \right)^{\frac{2}{n}}  } ,
\ee
with $c_{ln} = O(1)$, e.g.~$c_{32} = 0.92$, or more detailed

\be
c_{ln}  =  \frac{ l-1 }{ nl  \left( 2 + \frac{2nl}{l-2}  \right) }   \left[  l(l-1)  \right]^{ \frac{l}{l-2} }   \left[  \frac{ 16(l-1) }{ l(3l-2)^2 }  \right]^{ \frac{l}{l-2}+\frac{1}{n} }   \left[  \frac{ l^2(3l-2) }{ l-2 }  \right]^{ \frac{l}{2l-2} + \frac{l-2}{n(2l-2)} } .
\ee

From eq.~\eqref{eta0} the slow-roll predictions can be extracted. Note that for any choice of parameters, one finds $-\frac{1}{N_0} < \eta_0 < 0$, which translates (together with eq.~\eqref{epsilonEta} and \eqref{xiEta}) to general bounds on $n_s$, $r$ and $\alpha_s$ (assuming $N_0 \geq 50$):
\begin{align}
0.96  <  n_s  <  1, \\
\label{eq:approxR}
r  <  4 \times 10^{-5}, \\
\label{eq:approxAlphas}
-10^{-3}  <  \alpha_s   <  0.
\end{align}
$r$ and $\alpha_s$ are predicted to be unobservably small, and $n_s$ can be inside the WMAP $1\sigma$-bound of $n_s = 0.968 \pm 0.012$ (68\% CL) \cite{Komatsu:2010fb}. To get a sufficiently small $n_s$, the terms in the denominator of eq.~\eqref{eta0} should be small, so best agreement with data is achieved if $M$ and $\lambda$ are large\footnote{
Note that our approximations are only valid for $\kappa_S \gtrsim O(1)$, so eq.~\eqref{eta0} should not be applied for small $\kappa_S$.}. For typical parameter choices and $\lambda = O(1)$,  one usually finds $M \gtrsim O(M_{\text{GUT}})$.

\subsubsection{Constraint on $M$}
To find the range of allowed values for $M$, there is another condition to check: our treatment of supergravity as a low-energy effective field theory requires that the potential scaling parameter in eq.~\eqref{eqW} should be $\kappa \lesssim O(1)$.

We can determine $\kappa$ by normalizing our potential to the measured amplitude of scalar curvature perturbations $\Delta_{\mathcal{R}}^2 = (2.43 \pm 0.11) \times 10^{-9}$ \cite{Komatsu:2010fb}:
\begin{align}
& \Delta_{\mathcal{R}}^2 = \frac{1}{24\pi^2} \frac{ V(\phi_0) }{ \varepsilon_0 } \simeq \frac{1}{24\pi^2} \frac{ \kappa^2 M^4 }{ \varepsilon_0 }\\
\Rightarrow \quad & \kappa^2 \simeq 24\pi^2 \Delta_{\mathcal{R}}^2 \frac{ \varepsilon_0 }{ M^4 }  \lesssim  \frac{10^{-12}}{M^4},
\end{align}
where we have used $r = 16\,\varepsilon_0 < 4 \times 10^{-5}$. We see that $\kappa \lesssim O(1)$ is always satisfied for $M \gtrsim 10^{-3}$. With $\braket{H} = M^{2/l}$ and $l \geq 3$, we find that $\braket{H} \gtrsim O(10^{-2}) \simeq O(M_{\text{GUT}})$, so the phase transition is expected to happen at or above the GUT scale.

\section{Numerical results for the ($l=m=3$, $n=2$) case}
In this section we present our numerical results for the $(l=m,\,n)=(3,\,2)$ case. 
Up to this point we have included only the K\"ahler potential term 
$\Delta K = -\frac{1}{4} \kappa_S (S^{\dagger}S)^2$, in order to generate a heavy mass 
for the driving field $S$. However, we have checked that in a general expansion of the K\"ahler potential, one can choose coefficients such that in leading orders only the mass term for $S$  is generated. Furthermore, we have also checked numerically that the contribution of radiative corrections are suppressed in our model.

In our numerical results we consider $\kappa_S$ as the only extra parameter apart from $\kappa$, $\lambda$ and $M$. However, in the discussion of our numerical results we find it convenient to work with the parameter $\kappa \lambda$ instead of $\lambda$, as $\kappa \lambda$ is the actual coupling parameter of the nonrenormalizable term in the superpotential.

\begin{figure}[ht]
\centering \includegraphics[width=12 cm]{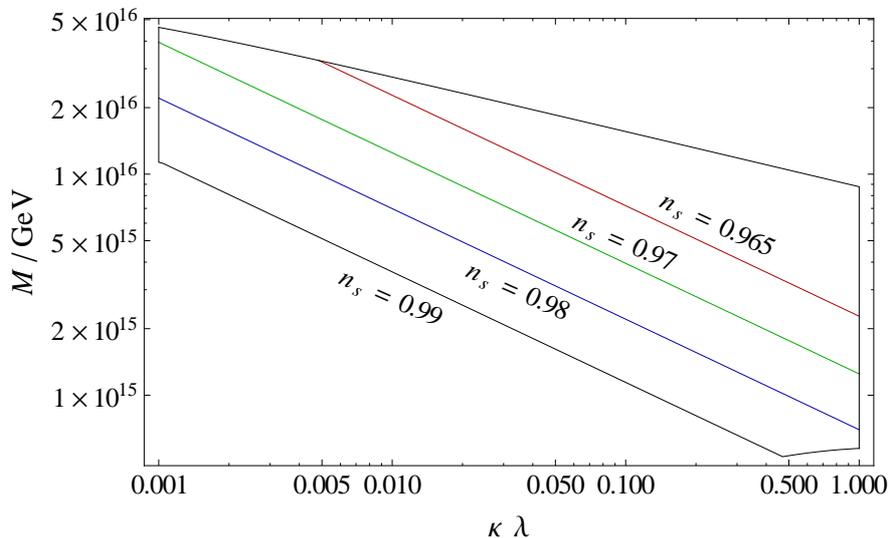}
\caption{The allowed range of $M$ and $\kappa \, \lambda$ for different values of $n_s$, 
with $10^{-3} \le \kappa \le 1$, $10^{-3} \le \kappa \lambda \le 1$, $N_0 = 50$ and $\kappa_S = 1$.
It is interesting to note that a large portion of the allowed range actually lies 
within the WMAP 1-$\sigma$ bound.}
\label{fig3}
\end{figure}

\begin{figure}[ht]
\centering \includegraphics[width=12 cm]{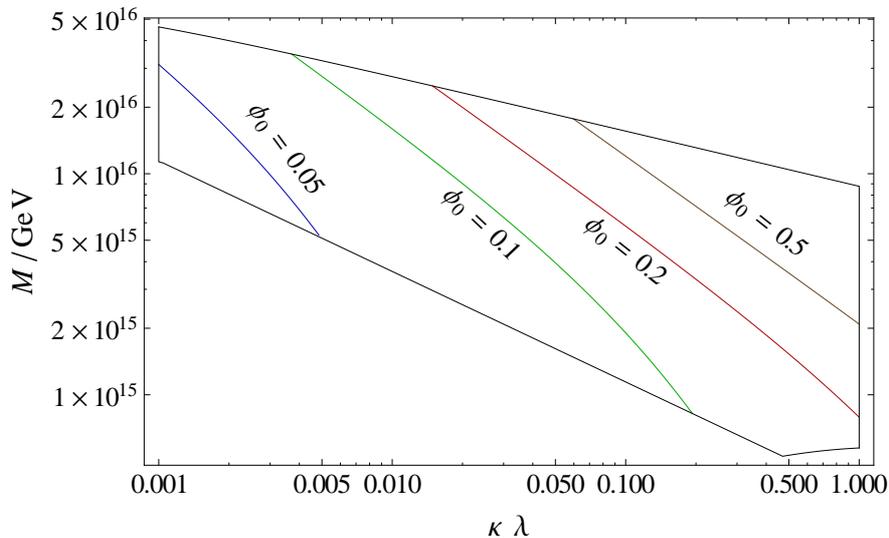}
\caption{The allowed range of $M$ and $\kappa \, \lambda$ for different values of $\phi_0$ (in Planck units), 
with $N_0 = 50$ and $\kappa_S = 1$.}
\label{fig4}
\end{figure}

\begin{figure}[th]
\centering \includegraphics[width=12 cm]{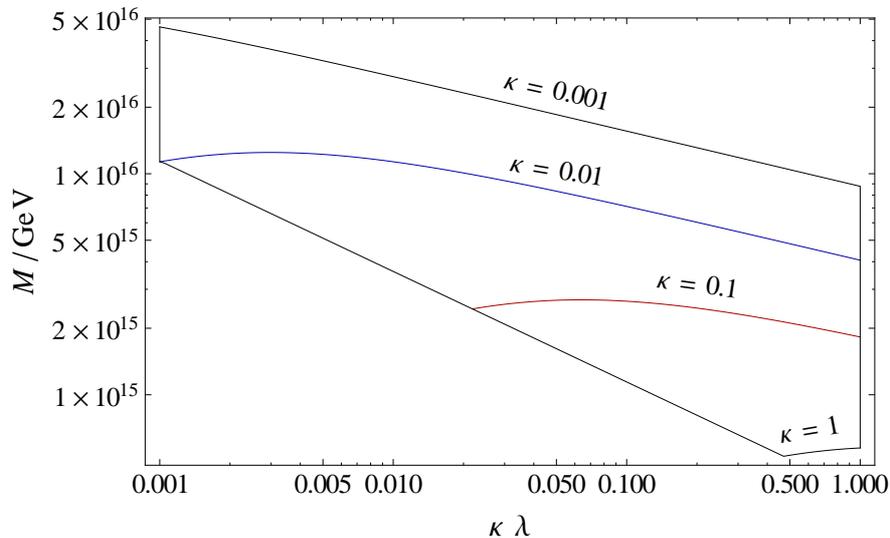}
\caption{The allowed range of $M$ and $\kappa \, \lambda$ for different values of $\kappa$, 
with $N_0 = 50$ and $\kappa_S = 1$.}
\label{fig5}
\end{figure}

The results of our calculations are displayed in Figs.~\ref{fig3}, \ref{fig4} and \ref{fig5}.
Here, we have set $\kappa_S = 1$ with $10^{-3} \le \kappa \le 1$, $10^{-3} \le \kappa \lambda \le 1$ and $N_0 = 50$.
To achieve better precision, we have also included the next-to-leading 
order corrections \cite{Stewart:1993bc,NeferSenoguz:2008nn} 
in the slow roll expansion for the quantities $n_s$, $r$, $\alpha_s$, and 
$\Delta_{\mathcal{R}}$. In all plots, the upper and lower boundaries in the $M$--$\kappa \lambda$ plane are given by
$0.001 \leq \kappa \leq 1$ and $n_s \leq 0.99$.

In Fig.~\ref{fig3}, we have plotted $M$ versus $\kappa \lambda$ for various
values of $n_s$. It is easy to see that our numerical results confirm the analytical approximations derived
in the previous section, more precisely we obtain $10^{-4} \lesssim M \lesssim 10^{-2}$
and $n_s \gtrsim 0.964$ within the WMAP 2-$\sigma$ bound. Note that the WMAP central value of $n_s = 0.968$ can only be achieved for sufficiently large $M \gtrsim 2 \times 10^{15}$ GeV and small $\kappa \lesssim 0.1$.
The behavior of $\phi_0$ and $\kappa$ is depicted in Figs.~\ref{fig4} and \ref{fig5}, respectively.

   It is important to note that $10^{16}$ GeV $ \lesssim \langle H \rangle  \lesssim 10^{17}$ GeV 
within the WMAP 2-$\sigma$ bound. 
This may allow us to realise this general class of inflationary models within GUTs. Moreover, as noted previously, the tensor to scalar ratio 
$ 5 \times 10^{-9} \lesssim r \lesssim 5 \times 10^{-6}$ and the running of the spectral index 
$-5 \times 10^{-5} \lesssim \alpha_s \lesssim -8 \times 10^{-4}$ turn out to be very small in these models.

\section{Summary and conclusions}

We have considered a new class of tribrid inflation models where the inflaton is a combination of Higgs and matter fields. Due to preselection of the vacuum during inflation the monopole problem is avoided. While at first sight the field dynamics looks similar to smooth inflation, it turns out that the tribrid models feature a waterfall, in contrast to smooth inflation.

We found that the tribrid superpotentials defined in eq.~\eqref{eqW} can feature suitable small-field trajectories only if $l \geq m > 2$. For the special case $l = m$ we analytically showed that the trajectory exhibits a waterfall instability and calculated the critical field value $\phi_c$ in eq.~\eqref{phic}. We also derived estimates for the spectrum of primordial density perturbations (eqs. \eqref{eta0}, \eqref{eq:approxR} and \eqref{eq:approxAlphas}). The numerical results agree well with these analytical estimates.

The predictions for the scalar spectral index $n_s \gtrsim 0.964$, the tensor to scalar ratio $r \lesssim 5 \times 10^{-6}$ and the running of the spectral index $-5 \times 10^{-5} \lesssim \alpha_s \lesssim -8 \times 10^{-4}$ are in good agreement with WMAP current observations and will be tested by the forthcoming data from the Planck satellite. With $ n_s \le 0.99$, $10^{-3} \le \kappa \le 1$ and $10^{-3} \le \kappa \lambda \le 1$, we obtain $10^{16}$ GeV $ \lesssim \langle H \rangle  \lesssim 10^{17}$ GeV.

The new class of models opens up new possibilities for realising inflation in close contact with particle physics, for instance with supersymmetric GUTs or with supersymmetric flavour models based on family symmetries.

\section*{Acknowledgments}
This work was supported by the Swiss National Science Foundation. S.~A.\ also acknowledges support by the DFG cluster of excellence ``Origin and Structure of the Universe''.

\end{document}